\documentclass[12pt]{article}
\usepackage{epsfig}
\usepackage[normal] {subfigure}
\usepackage{calc}

\def\beq{\begin{equation}}
\def\eeq#1{\label{#1}\end{equation}}
\def\eeqn{\end{equation}}

\def\beqa{\begin{eqnarray}}
\def\eeqa#1{\label{#1}\end{eqnarray}}
\def\eeqan{\end{eqnarray}}

\let\bar=\overbar

\def\Dslash{\not{\hbox{\kern-4pt $D$}}}
\def\dslash{\not{\hbox{\kern-2pt $\del$}}}

\def\msb{{\bar{\ssstyle M \kern -1pt S}}}

\def\Title#1{\begin{center} {\Large {\bf #1} } \end{center}}
\begin{document}
\Title{Measurements of Electroweak Top\vspace{1mm}\\Production at the LHC}
\bigskip\bigskip


\begin{raggedright}  
{\it Aniello Spiezia\index{Spiezia, A.}\\
on behalf of the ATLAS and CMS Collaborations\\
Dipartimento di Fisica\\
INFN and Universit\'a degli Studi di Perugia\\
06123 Perugia, ITALY}\vspace{3mm}\\
\end{raggedright}
\begin{center}
Proceedings of CKM 2012, the 7th International Workshop on the CKM Unitarity Triangle, University of Cincinnati, USA, 28 September - 2 October 2012
\end{center}\vspace{-3mm}
\bigskip\bigskip

\section{Introduction}
\begin{figure}[h]
\begin{center}
\includegraphics[scale=0.6]{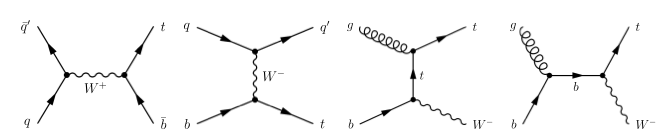}
\caption{Top-quark production modes at LHC: s-channel exchange of a virtual W boson (left), the t-channel exchange of a virtual W boson (middle-left) and the associated production of a top quark and a W boson (tW) (middle-right and right).} \label{top_production}
\end{center}
\end{figure}

\vspace{-2mm}
The production of a single top quark at LHC may occur through charged-current electroweak interactions, i.e. via the three different processes reported in figure \ref{top_production}: the s-channel exchange of a virtual W boson, the t-channel exchange of a virtual W boson and the associated production of a top quark and a W boson (tW). 
At LHC, at a center-of-mass energy of 7 TeV, the cross sections for these processes are predicted to be:  $\sigma_{t\textrm{-}channel} = 64.6^{+2.7}_{-2.0} \; pb$, $\sigma_{tW} = 15.7 \pm 1.1 \; pb$ and $\sigma_{s\textrm{-}channel}  = 4.6 \pm 0.2 \; pb$ \cite{ATLAS7TeV_tch_v1}. 
The importance of these processes lies in the fact that they can be used to test the predictions of the standard model (SM) of particle physics and to search for new phenomena. Furthermore they can give a direct measurement of the magnitude of the Cabibbo-Kobayashi-Maskawa (CKM) matrix element $|V_{tb}|$.
\\
In the following a review of the measurements of the single-top production made both at ATLAS and CMS will be shown. The three processes will be described separately in the next three sections.

\section{t-channel production}\label{tchannel}
\begin{figure}[h]\vspace{-3mm}
\begin{center}
\subfigure{\includegraphics[scale=0.44]{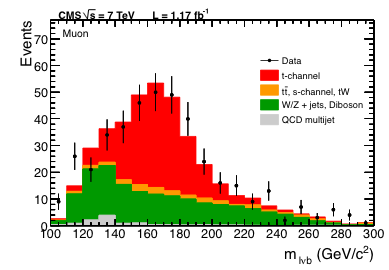}}\hspace{5mm}
\subfigure{\includegraphics[scale=0.14]{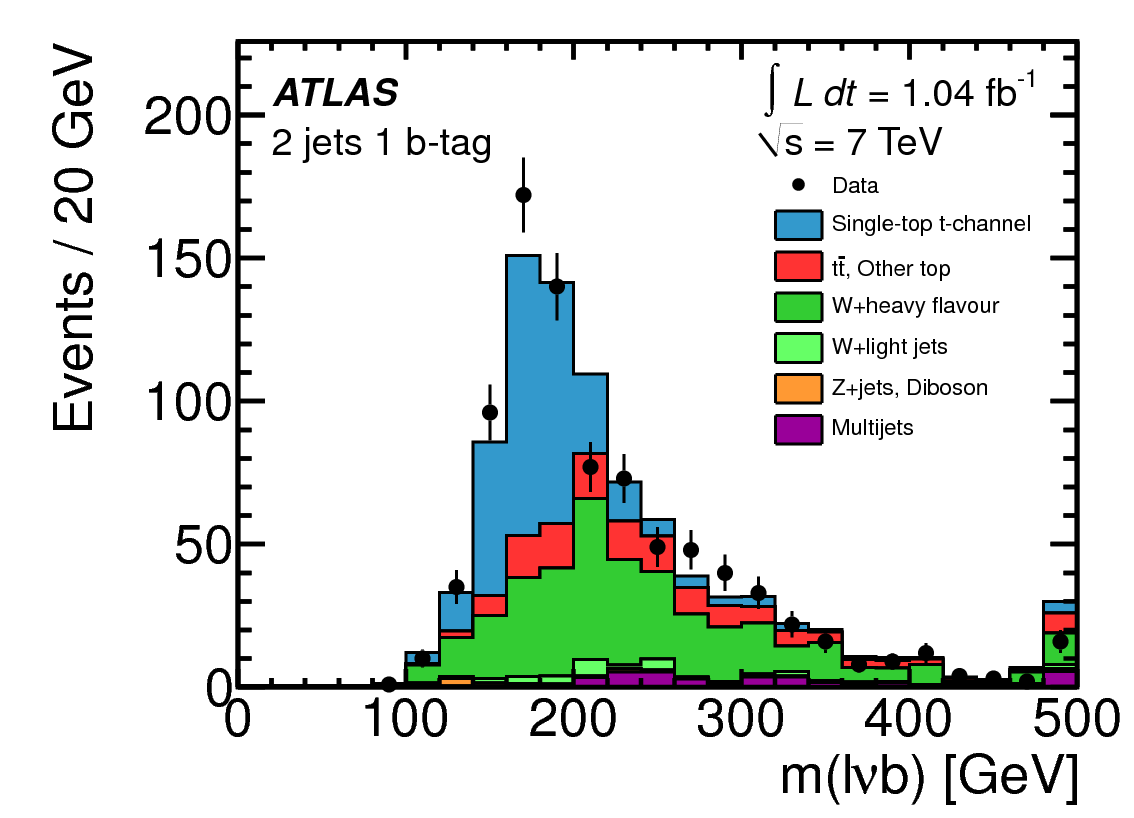}}
\end{center}\vspace{-7mm}
\caption{Reconstructed top-quark mass for CMS \cite{CMS7TeV_tch} (left) and ATLAS \cite{ATLAS7TeV_tch_v1} (right) at 7 TeV.} \label{M_lnb}
\end{figure}

\vspace{-2mm}
Both ATLAS and CMS studied the t-channel production of the single-top quark at 7 TeV and at 8 TeV, using events with leptonically decay of the W boson: $t \rightarrow bW \rightarrow b\ell\nu$, where $\ell$ can be a muon or an electron. As can be seen in figure \ref{top_production} these events are characterized by a final state with one jet, likely a light jet, and the products of the top-quark, i.e. a muon or an electron, a b-quark and a neutrino that can be seen as transverse missing energy in the event. Using this final state, ATLAS and CMS studied the t-channel using different analysis.\\
For 7 TeV data, CMS developed three different analyses \cite{CMS7TeV_tch} on 1.17 (1.56) $fb^{-1}$ for muon (electron): one based on simple cuts and on the study of the $|\eta_{j'}|$ distribution, namely the pseudo-rapidity of the light jet, and two more precise, based on multivariate analysis (MVA) techniques. 
The results of the three analyses are combined using the BLUE (Best Linear Unbiased Estimator) method giving $\sigma_{t\textrm{-}channel} = 67.2 \pm 3.7 (stat) \pm 3.0 (syst)  \pm 3.5 (theor) \pm 1.5 (lum) \; pb$. Assuming $|V_{td}|$ and $|V_{ts}|$ much smaller than $|V_{tb}|$, this can be measured from $|V_{tb}| = \sqrt{\sigma_{exp}/\sigma_{th}}$ where $\sigma_{th}$ is the SM prediction with $|V_{tb}|=1$, so that $|f_{L_V}$ $V_{tb}| = 1.020 \pm 0.046 (exp) \pm 0.017 (theor)$ (where $f_{L_V}$, introduced for \textit{anomalous coupling}, is one in the SM), and if $|V_{tb}|$ is requested to be less than one, 
\begin{figure}[h]\vspace{-2mm}
\begin{center}
\subfigure{\includegraphics[scale=0.36]{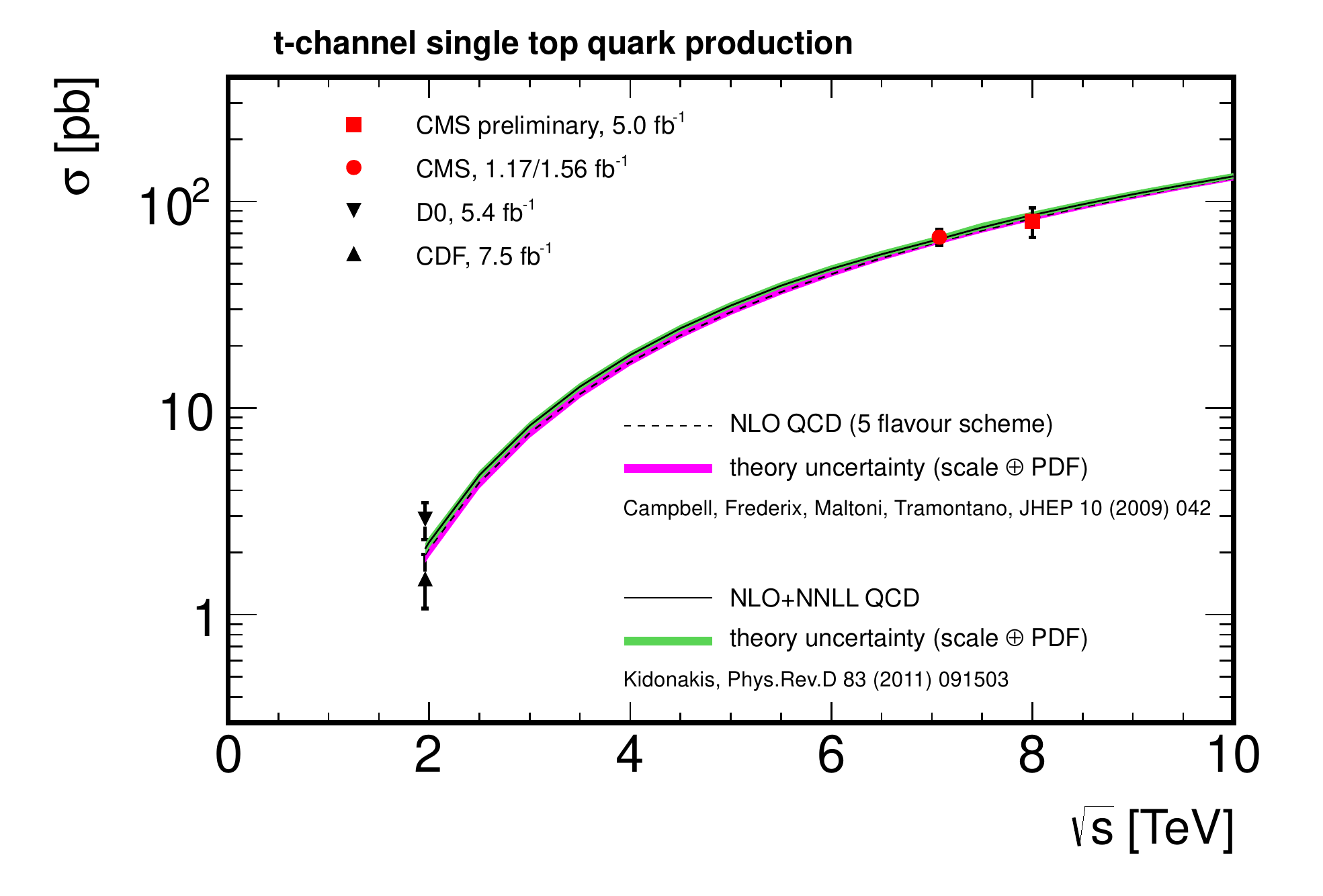}}\hspace{3mm}
\subfigure{\includegraphics[scale=0.15]{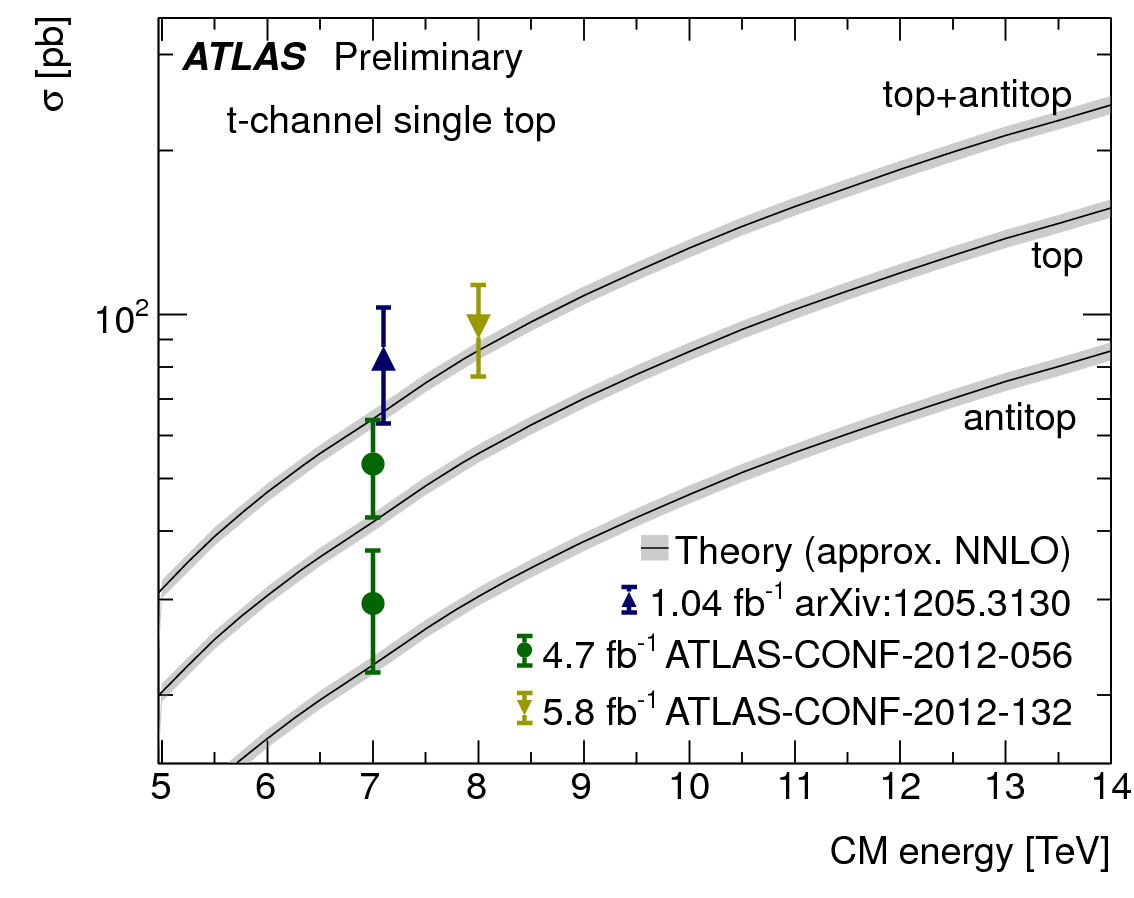}}
\end{center}\vspace{-5mm}
\caption{Cross section measurements and SM expectation for the t-channel single top-quark measurement at 7 and 8 TeV for CMS \cite{CMS8TeV_tch} (left) and ATLAS \cite{ATLAS8TeV_tch} (right).} \label{tch_CMS}
\end{figure}
 the measured limit is $0.92 < |V_{tb}| < 1$, at 95\% CL.\\
The CMS 8 TeV measurement \cite{CMS8TeV_tch} follows the $|\eta_{j'}|$ analysis and focalizes on the muon channel only with a statistics of $5.0\; fb^{-1}$. The 
measured cross section is $\sigma_{t\textrm{-}channel} = 80.1 \pm 5.7 (stat) \pm 11.0 (syst)  \pm 4.0 (lum) \; pb$. In figure \ref{tch_CMS} (left), it can be seen the agreement between these two measurements with the SM expectations.\\
ATLAS single top-quark search in the t-channel at 7 TeV is based on $1.04\; fb^{-1}$ \cite{ATLAS7TeV_tch_v1}. The measurement is performed using a neural network and is cross-checked with a cut-based method. 
In order to separate signal from background events, several variables are combined into one discriminant by employing a neural network method: 12 variables are used for the 2 jets tagged sample (between them the most discriminant are $m(\ell\nu b)$ and $|\eta(j_u)|$) and 18 variables for the 3 jets tagged sample ($m(j_1,j_2)$, $m(\ell\nu b)$, etc.). A simultaneously binned maximum-likelihood fit to the NN outputs of the two samples is performed, 
giving the measurement of the cross section $\sigma_{t\textrm{-}channel} = 83 \pm 4 (stat)  ^{+20}_{-19} (syst) \; pb$, while the measured CKM matrix element is $|V_{tb}| = 1.13 ^{+0.14}_{-0.13} (stat) \pm 0.02 (theor)$ and restricting to the region below one, the limit is $0.75 < |V_{tb}| < 1$ at 95\% CL.\\ 
ATLAS also provides a measurement of the ratio $R_{t} = \sigma_{t\textrm{-}channel}(t)/\sigma_{t\textrm{-}channel}(\bar{t})$, i.e. of the cross sections of the top and antitop quark productions \cite{ATLAS7TeV_tch_v2}. Since the up-quark density of the proton is about twice the down-quark density, $\sigma_{t\textrm{-}channel}(t)$ is expected to be about twice $\sigma_{t\textrm{-}channel}(\bar{t})$ and the ratio $R_{t}$ is expected to be near two. ATLAS measured $R_t = 1.81 \pm 0.10 (stat) ^{+0.21}_{-0.20} (syst)$, in agreement with the SM prediction.\\
The ATLAS 8 TeV measurement \cite{ATLAS8TeV_tch} follows the same NN analysis strategy. 
The measured cross section at 8 TeV, obtained analyzing 5.8 $fb^{-1}$, is $\sigma_{t\textrm{-}channel} = 95 \pm 2 (stat) \pm 18 (syst)  \; pb$. In figure \ref{tch_CMS} (right) it is shown the agreement between the measured cross sections at 7 and 8 TeV with the SM prediction for the ATLAS analysis.\\
In figure \ref{M_lnb}, a comparison between background prediction and data observation is shown for the reconstructed top-quark mass for CMS (left) and ATLAS (right) at 7 TeV. 

\section{tW channel production}\label{tWchannel}\vspace{-2mm}
The tW production is inaccessible at Tevatron and it was observed for the first time at ATLAS and CMS. Looking for events in which the two W bosons (the one produced in association with the top and the one coming from the top quark, see figure \ref{top_production}) decay leptonically, as it is done at both ATLAS and CMS, the final state is characterized by the presence of two isolated leptons with opposite charge ($ee$, $e\mu$ and $\mu\mu$ in the following), one b-jet and two neutrinos.\\
In CMS \cite{CMS7TeV_tW}, 
a multivariate analysis based on boosted decision trees (BDT) is used, in which four variables are chosen to train the BDT: $H_T$, defined as the scalar sum of the transverse momenta of the leptons, jet and $E_T^{miss}$; the $p_T$ of the system formed by the two leptons, the jet and the $E_T^{miss}$; the $p_T$ of the leading jet and finally the angular distance between the direction associated to the $E_T^{miss}$ and the closest of the two selected leptons. 
The measured cross section, CKM element matrix and the limit on it are: $\sigma_{tW} = 16^{+5}_{-4} \; pb$, $|V_{tb}| = 1.01^{+0.16}_{-0.13} (exp) ^{+0.03}_{-0.04} (th)$ and $0.79 < |V_{tb}| < 1$ at 90 \% CL.\\
The signal selection for the ATLAS analysis \cite{ATLAS7TeV_tW} is very similar to the one described for the CMS analysis and the main difference is that no b-tagging is applied on the jets. 
Then a multivariate approach based on boosted decision trees (BDT) is used taking as input 22 variables and the measurement of the cross section is obtained from a fit of the three BDT output distributions (events with one jet, events with two jets and events with more than two jets), giving $\sigma_{tW} = 16.8 \pm 2.9 (stat) \pm 4.9 (syst) \; pb$ and $|V_{tb}| = 1.03 ^{+0.16} _{-0.19}$.

\section{s-channel production}\label{schannel}\vspace{-2mm}
The s-channel production mode has the lowest cross section between the production modes of the single top quark. The search for it is interesting since it is sensitive to several models of new physics, like $W'$ bosons or charged Higgs boson. Only ATLAS has presented a search of this channel, putting a limit on its cross section \cite{ATLAS7TeV_sch}. As can be seen from figure \ref{top_production}, the final state is characterized by one lepton (muon or electron in this analysis), a neutrino and two b-jets. 
The measurement of the s-channel single top-quark signal is challenging since it is overwhelmed by large background processes, therefore after the preselection just described, 
\begin{figure}[ht]
\begin{center}
\subfigure{\includegraphics[scale=0.13]{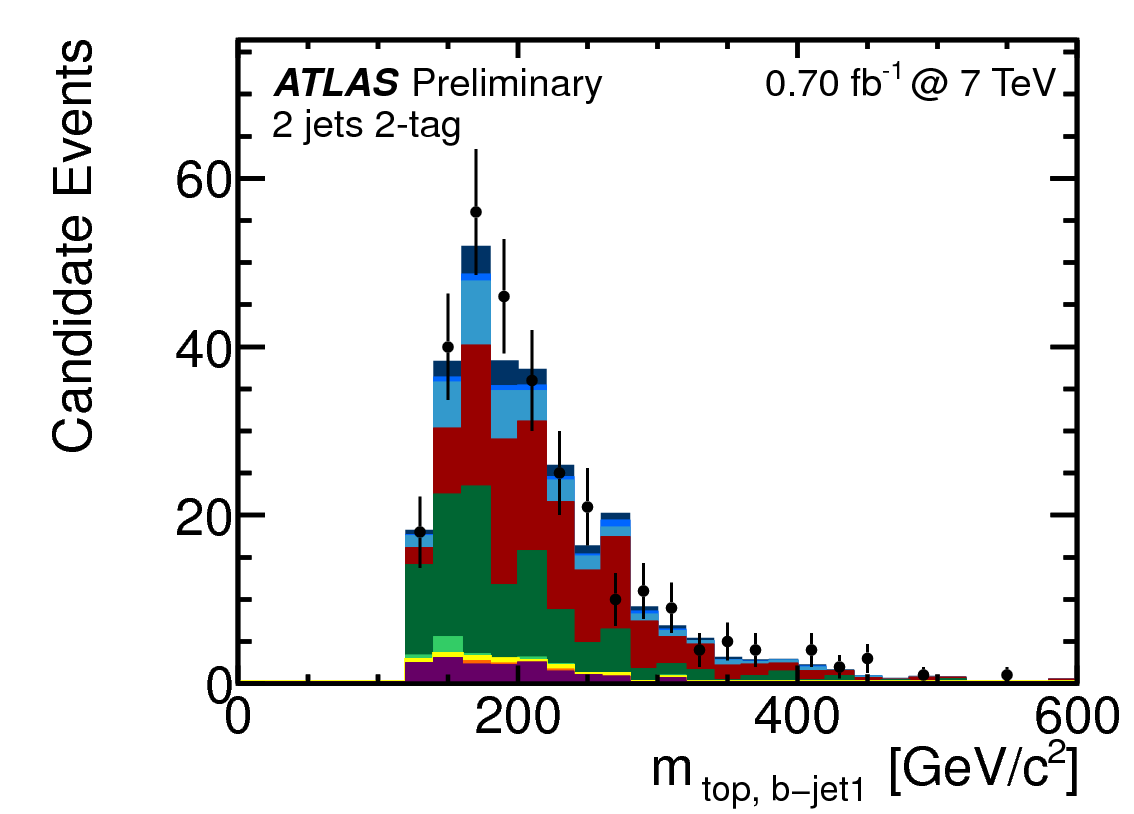}}
\subfigure{\includegraphics[scale=0.13]{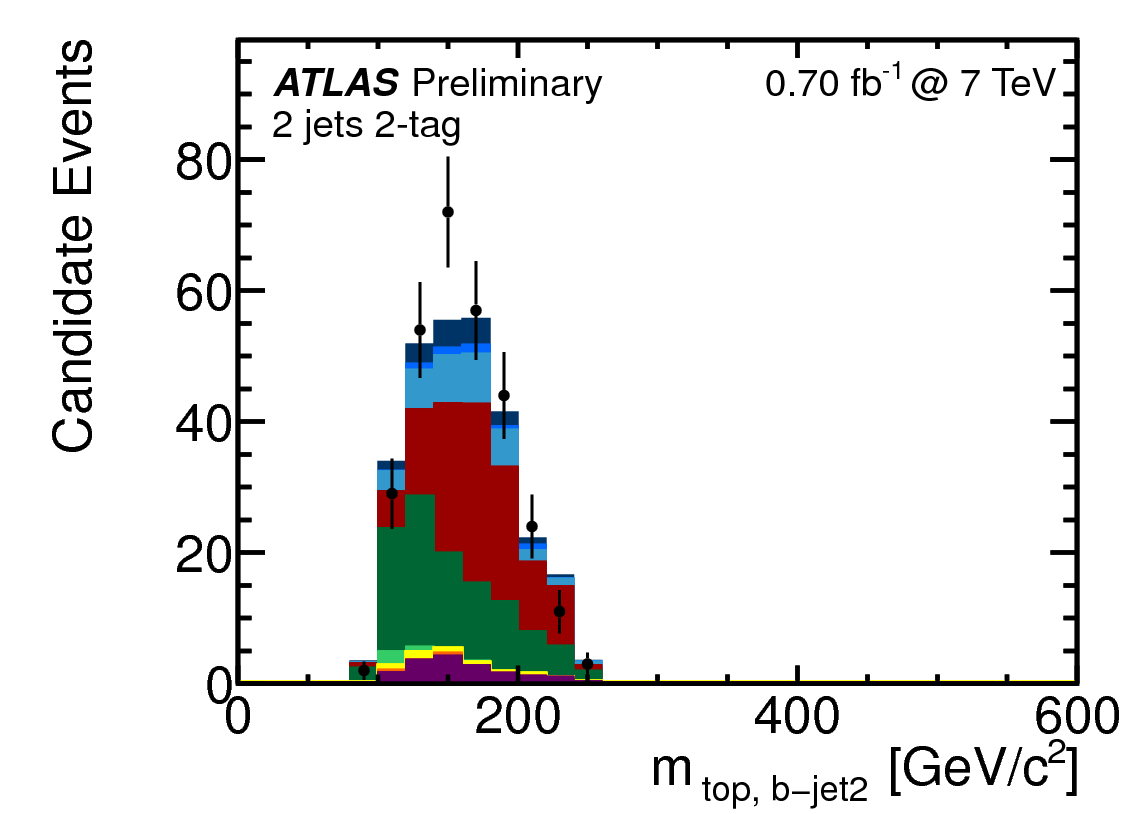}}\hspace{5mm}
\raisebox{10mm}{\subfigure{\includegraphics[scale=0.08]{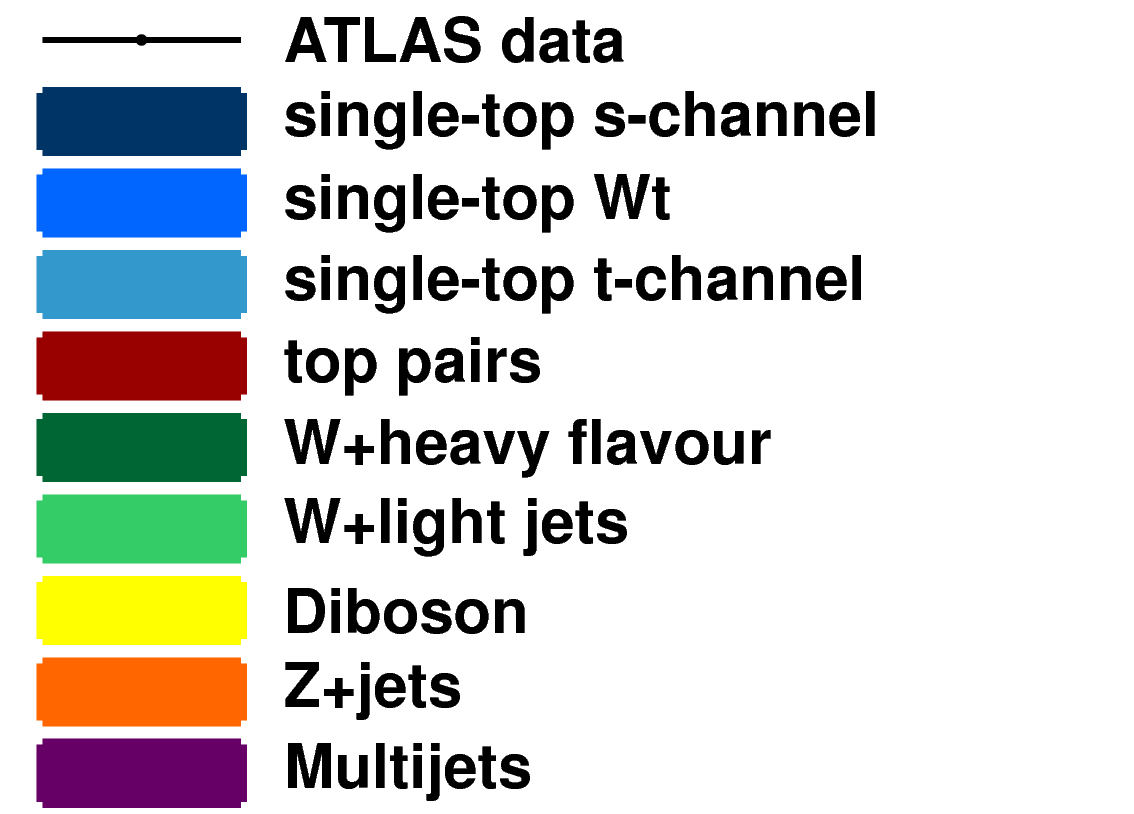}}}
\end{center}\vspace{-8mm}
\caption{Distribution of the invariant mass of the leading (left) or subleading (right) jet, the lepton and the neutrino for electron and muon two-jets events after the tighter cuts are applied \cite{ATLAS7TeV_sch}.} \label{s_ch}
\end{figure}
tighter requirements are applied in order to isolate the s-channel signal. Finally, on the 296 events that pass those cuts, $16\pm6$ signal events are expected to be in the selected sample. Due to that, only a limit on the cross section can be put: $\sigma (s\textrm{-}channel) < 26.5$ pb at 95\% CL. In figure \ref{s_ch} the agreement between prediction and observation for the invariant mass between the leading or the subleading jet, the lepton and the neutrino is shown.

\section{Conclusion}\label{conclusion}\vspace{-2mm}
An overview of the measurements made at the LHC for the single top-quark production in the three production mechanisms described in figure \ref{top_production} has been presented. The measurements of CMS and ATLAS show a good agreement between them and with the SM prediction and are summarized in table \ref{measurements}.
\begin{table}[h]
\begin{center} 
\begin{tabular}{|c|c|c|c|}
\hline
 & \textbf{Theor.} & \textbf{CMS} & \textbf{ATLAS}\\\hline
\textbf{t-channel (7 TeV)} & $64.6^{+2.7}_{-2.0} \; pb$ \cite{ATLAS7TeV_tch_v1}& $67.2 \pm 6.1 \; pb$ & $82.7 \pm 18.1 \; pb$ \\\hline
\textbf{t-channel (8 TeV)} & $87.8^{+3.4}_{-1.9} \; pb$ \cite{ATLAS7TeV_tch_v1}& $80.1 \pm 13.0 \; pb$ & $95.1 \pm 18.1 \; pb$ \\\hline
\textbf{tW (7 TeV)} & $15.7 \pm 1.1 \; pb$ \cite{ATLAS7TeV_tch_v1}&  $16^{+5}_{-4} \; pb$ & $16.8 \pm 5.7 \; pb$ \\\hline
\textbf{s-channel (7 TeV)} & $4.6 \pm 0.2 \; pb$ \cite{ATLAS7TeV_tch_v1}&  - & $ < 26.5 \; pb$\\\hline
\end{tabular}
\end{center} \vspace{-3mm}
\caption{Predicted and measured cross sections for the three single top-quark production modes at CMS and ATLAS.} \label{measurements}
\end{table}
\vspace{-5mm}


\end{document}